%% file: main.tex
\documentstyle[12pt,psfig]{article} 
\begin{document}


\begin{center}
{\Large \bf
 Statistical Multifragmentation in Thermodynamical Limit:\\
An Exact Solution for Phase Transitions
}

\vspace{1.0cm}

{\bf \underline{K.A. Bugaev$^{1,2}$,} M.I. Gorenstein$^{1,2}$,
I.N. Mishustin$^{1,3,4}$} and\\ {\bf W. Greiner$^{1}$}

\vspace{1.0cm}
\noindent

\begin{minipage}[b]{11.0cm}
$^1$ Institut f\"ur Theoretische Physik,
Universit\"at Frankfurt, Germany\\
$^2$ Bogolyubov Institute for Theoretical Physics,
Kyiv, Ukraine\\
$^3$ Kurchatov Institute, Russian Research Center,
Moscow, Russia\\
$^4$ Niels Bohr Institute, University of Copenhagen, Denmark
\end{minipage}

\end{center}

\vspace{1cm}

\begin{abstract}
An exact analytical solution of the 
statistical  multifragmentation model is found in thermodynamic limit.
Excluded volume effects are taken into account in the thermodynamically 
self-consistent way.
The model exhibits a 1-st order 
phase transition of the liquid-gas type.
An extension of the model including the Fisher's term is also studied.
The possibility of the second order phase transition at or above the critical 
point is discussed.
The mixed phase region of the phase diagram, where
the gas of nuclear fragments coexists with the infinite
liquid condensate, is unambiguously identified.
The peculiar thermodynamic properties of the model near the boundary
between the mixed phase
and the pure gaseous phase are studied.
The results for the caloric curve and specific heat are presented
and a physical picture 
of the nuclear liquid-gas phase transition is clarified.
\end{abstract}

\vspace*{0.5cm}

\noindent
\hspace*{1.0cm}\begin{minipage}[t]{11.cm}
mixed phase thermodynamics
\end{minipage}

\vspace*{0.3cm}

\noindent
\hspace*{1.0cm}

\newpage

Nuclear multifragmentation is one of 
the most interesting and widely discussed phenomena in
intermediate energy nuclear reactions.
The statistical multifragmentation model (SMM) 
(see \cite{Bo:95,Gr:97} and references therein)
was recently applied
to study 
the liquid-gas phase transition
in nuclear matter
\cite{Ch:95,Gu:98,Gu:99,Gu:00}. Numerical calculations
within the canonical ensemble exhibited
many intriguing peculiarities of the finite multifragment
systems. 
However, the investigation of the system's
behavior in the thermodynamic limit was still missing.
Therefore, there was no rigorous proof of the phase
transition existence, and the phase diagram structure
of the SMM remained unclear.
Previous numerical studies for the finite nuclear systems
(the canonical and microcanonical ensembles) led to 
unjustified
(and sometimes wrong) statements concerning  the nuclear liquid-gas phase
transition in the thermodynamic limit.
In our recent paper \cite{Bu:00} an exact analytical solution
of the SMM was found within the grand canonical ensemble
which naturally allowed to study the thermodynamic limit. 
The self-consistent treatment
of the excluded volume effects
was an important part of this study. 
In this letter we investigate
the peculiar thermodynamic properties near the boundary
between the mixed phase
and the pure gaseous phase. 
New results for the caloric curve and the specific heat are presented
and a physical picture
of the nuclear liquid-gas phase transition in SMM is clarified.
This physical picture differs from the one advocated in the previous 
numerical studies.

In the SMM the states of the system  are specified by the multiplicity
sets  $\{n_k\}$
($n_k=0,1,2,...$) of $k$-nucleon fragments.
The partition function of a single fragment with $k$ nucleons is
\cite{Bo:95}:
$
\omega_k =V\left(m T k/2\pi\right)^{3/2}z_k~
$,
where $k=1,2,...,A$ ($A$ is the total number of nucleons
in the system). $V$ and $T$ are, respectively, the  volume
and the temperature of the system,
$m$ is the
nucleon mass.
The first two factors in $\omega_k$ originate
from the
non-relativistic thermal motion 
and the last factor,
 $z_k$, represents the intrinsic partition function of the
$k$-fragment.
For \mbox{$k=1$} (nucleon) we take $z_1=4$ 
(4 internal spin-isospin states) 
and for fragments with $k>1$ we use the expression motivated by the
liquid drop model (see details in \mbox{Ref. \cite{Bo:95}):} 
$
z_k=\exp(-f_k/T),
$ with the fragment free energy 
\begin{equation}\label{fk}
f_k~ = ~- [W_{\rm o}~+~
T^2/\epsilon_{\rm o}~]k~+~\sigma (T)~ k^{2/3}~+~\tau~ T\ln k~.
\end{equation}
Here $W_{\rm o}=16$~MeV is the bulk binding energy per nucleon,
$T^2/\epsilon_{\rm o}$ is the contribution of 
the excited states taken in the Fermi-gas
approximation ($\epsilon_{\rm o}=16$~MeV) and $\sigma (T)$ is the
temperature dependent surface tension which is parameterized 
in the following form:
\begin{equation}\label{Sig}
\sigma (T)=\sigma_{\rm o}
[(T_c^2~-~T^2)/(T_c^2~+~T^2)]^{5/4},
\end{equation}
with $\sigma_{\rm o}=18$~MeV and $T_c=18$~MeV ($\sigma=0$
at $T \ge T_c$). The last Fisher's term in Eq.~(\ref{fk}) with
dimensionless parameter
$\tau$ is introduced for generality.
The canonical partition function (CPF) of the ensemble of nuclear
fragments 
has the following form:
\begin{equation} \label{Zc}
Z^{id}_A(V,T)~=~\sum_{\{n_k\}}~\prod_{k=1}^{A}~\frac{\omega_k^{n_k}}{n_k!}~
\delta(A-\sum_k kn_k)~.
\end{equation}
The model defined by Eqs.(\ref{fk},\ref{Zc}) with $\tau=0$
was studied numerically in Refs.~\cite{Ch:95,Gu:98,Gu:99,Gu:00}.
This is a simplified version of the SMM since the symmetry-energy  and
Coulomb
contributions are neglected. However, its investigation appears to be 
very important for understanding the physics of multifragmentation.

In Eq. (\ref{Zc}) the nuclear fragments are treated as point-like objects.
However, these fragments have non-zero proper volumes and
they should not overlap
in the coordinate space. 
In the 
Van der Waals excluded volume 
approximation 
this is achieved
by replacing
the total volume $V$
in Eq. (\ref{Zc}) by the free (available) volume 
$V_f\equiv V-b\sum_k kn_k$, where
$b=1/\rho_{{\rm o}}$
($\rho_{{\rm o}}=0.16$~fm$^{-3}$ is the normal nuclear density).  
Therefore, the corrected CPF becomes:
$
Z_A(V,T)=Z^{id}_A(V-bA,T)
$.

The calculation of $Z_A(V,T)$
is difficult because of the constraint $\sum_k kn_k =A$.
This difficulty can be partly avoided by calculating the grand canonical
partition function:
\begin{equation} 
{\cal Z}(V,T,\mu)~\equiv~\sum_{A=0}^{\infty}
\exp\left(\mu A/T \right)
~Z_A(V,T)~\Theta (V-bA) \label{Zgc}~, 
\end{equation}
where the chemical potential $\mu$ is introduced.
The calculation of ${\cal Z}$  is still rather
difficult. The summation over the sets $\{n_k\}$
in $Z_A$ cannot be performed analytically because of
the additional \mbox{$A$-dependence}
in the free volume $V_f$ and the restriction
$V_f>0 $.
The problem can be solved 
by introducing the so-called
isobaric partition function (IPF) which is calculated
in a straightforward way (see details 
in Refs.~\cite{Bu:00, Go:81, Bu:00B, Bu:00C}):
\begin{equation} \label{Zs}
\hat{\cal Z}(s,T,\mu) ~ \equiv ~ \int_0^{\infty}dV~\exp(-sV)
~{\cal Z}(V,T,\mu)  
~ = ~ \frac{1}{s~-~{\cal F}(s,T,\mu)}~,
\end{equation}
where
\begin{eqnarray}
{\cal F}(s,T,\mu) &=&
\left( \frac{mT }{2\pi}\right)^{3/2} 
\left[z_1 \exp\left(\frac{\mu-sbT}{T}\right)  
\right.
\nonumber \\
&+& \sum_{k=2}^{\infty} 
\left.
k^{3/2 - \tau} \exp\left(
\frac{(\nu - sbT)k -
\sigma k^{2/3}}{T}\right)\right]\,\,, 
\label{Fs}
\end{eqnarray}
with 
$
\nu \equiv \mu + W_{\rm o}+T^2/\epsilon_{\rm o}
$.
In the thermodynamic limit $V\rightarrow \infty$ the pressure
of the system
is defined by the farthest-right singularity, $s^*(T,\mu)$, of  
the IPF $\hat{\cal Z}(s,T,\mu)$
\begin{equation}\label{ptmu}
p(T,\mu)~\equiv~ T~\lim_{V\rightarrow \infty}\frac{\ln~{\cal Z}(V,T,\mu)}
{V}~=~T~s^*(T,\mu)~.
\end{equation}
The study of the 
system's behavior in the thermodynamic limit
is therefore reduced to the investigation of
the singularities \mbox{of $\hat{\cal Z}$.}

The IPF (\ref{Zs})  has two types of singularities:
\mbox{1) the simple} pole singularity
defined by the following equation
\mbox{$s_g(T,\mu)= {\cal F}(s_g,T,\mu)~$;}
2) the singularity  of the function ${\cal F}$ 
 itself at the point $s_l(T,\mu)=\nu/Tb$ where the coefficient 
in linear over $k$ terms of the exponent in Eq.~(\ref{Fs}) 
is equal to zero.

The simple pole singularity corresponds to the gaseous phase 
where pressure $p_g\equiv Ts_g$ is determined by the 
following transcendental
equation:
$
p_g(T,\mu)=T{\cal F}(p_g/T,T,\mu)
$.
The singularity $s_l(T,\mu)$ of the function ${\cal F}$
defines the liquid pressure:
$
p_l(T,\mu)\equiv Ts_l(T,\mu)=
{\nu}/{b}.
$
Here the liquid is represented by an infinite fragment
(condensate) with $k=\infty$.

In the region of the $(T,\mu)$-plane where $\nu < bp_g(T,\mu)$ the
gaseous phase
is realized ($p_g > p_l$), while  the liquid phase
dominates at $\nu > b p_g(T,\mu)$. The liquid-gas phase transition
occurs when  the two singularities coincide,
i.e. $s_g(T,\mu)=s_l(T,\mu)$.
As ${\cal F}$ in Eq. (\ref{Fs}) 
is a monotonously decreasing
function of $s$ 
the necessary condition for the phase
transition is that the function
${\cal F}$ is finite in its singular
point $s_l$. 
At $\tau =0$ this condition requires $\sigma(T) >0$ and, therefore,
$T<T_c$.
Otherwise, ${\cal F}(s_l,T,\mu)=\infty$ and the system
is always in the gaseous phase as $s_g>s_l$. 
As one can see from Eq.(\ref{Fs}) the convergence properties of
${\cal F}(s,T,\mu)$ depend significantly on the Fisher's exponent
$\tau$ in the vicinity of the critical point where the surface
term vanishes.
In what follows we mainly concentrate 
on the  case  $\tau =0$. Other possibilities
which appear at $\tau >0$ are discussed shortly. 
Their detail study can be found in Ref.~\cite{Bu:00}.
Here we only note that Eqs. (\ref{Zs}, \ref{Fs}) represent an exact 
solution of the Fisher's droplet model \cite{Fisher:67} 
where additionally the effects of excluded volume are
incorporated.

The baryonic density $\rho$ in the liquid and gaseous phases
is given by the following formulae, 
respectively: 
\begin{equation}\label{rho}
 \rho_l  \equiv  
\left(\partial  p_l/\partial \mu\right)_{T}
=1/b~,~~~~
\rho_g \equiv
\left(\partial  p_g/\partial \mu\right)_{T}=
\rho_{id}/( 1 + b \rho_{id} )~,
\end{equation}
where the function $ \rho_{id}$ is the density of point-like
nuclear fragments with shifted, 
$
\mu \rightarrow \mu -bp_g
$,
chemical potential:
\begin{eqnarray}
\rho_{id}(T,\mu) &=& \left( \frac{mT }{2\pi}\right)^{3/2} 
\left[ z_1 \exp\left(\frac{\mu-bp_g}{T}\right) 
\right.
\nonumber \\
&+& \left. \sum_{k=2}^{\infty} 
k^{5/2} \exp\left(
\frac{(\nu - bp_g)k -
\sigma k^{2/3}}{T} \right) \right]~.
\label{rhoid}
\end{eqnarray}

At $T<T_c$ the system undergoes a 1-st order phase transition
across the line $\mu^*=\mu^*(T)$ defined by
the condition of coinciding singularities:
$ s_l=s_g,  $ i.e., $p_l = p_g$. 
The phase transition line 
$\mu^*(T)$
in the $(T,\mu)$-plane
corresponds to the mixed liquid and gas
states. This line 
is transformed into
the finite mixed-phase region in the $(T,\rho)$-plane
shown in Fig. 1. 
The baryonic density 
in the mixed phase
is a superposition of the liquid and gas baryonic densities:
$
\rho=\lambda\rho_l+(1-\lambda)\rho_g~,
$
where $\lambda$ ($0<\lambda <1$) is the fraction of the system's volume
occupied by the liquid  inside the mixed phase.
Similar linear combinations are also valid for the entropy density $s$
and the energy density $\varepsilon$ 
with $(i=l,g)$
$s_i=\left(\partial p_i/\partial T\right)_{\mu}$\,,
$\varepsilon_i=T \left(\partial p_i/\partial T \right)_{\mu} +
\mu \left(\partial p_i/\partial \mu\right)_{T} -p_i$.

Inside the mixed phase at constant density $\rho$ the
parameter $\lambda$ has a specific temperature dependence
shown in Fig. 2:
from an approximately
constant value $\rho/\rho_{\rm{o}}$ at small $T$ the function 
$\lambda(T)$ drops to zero in a narrow
vicinity of the boundary separating the  mixed phase and 
the pure gaseous phase.
This corresponds to a fast change of the configurations from
the state which is  dominated by one infinite liquid fragment to 
the gaseous multifragment configurations. It happens inside the
mixed phase  without
discontinuities in the thermodynamical functions.

An abrupt decrease of $\lambda(T)$ near this boundary
causes a strong
increase of the energy density as a function of temperature.
This is evident from Fig.~3 which shows the caloric curves at different
baryonic densities. One can clearly see a 
leveling of temperature at energies per nucleon between 10 
and 20 MeV.
As a consequence this  
leads to a sharp peak 
in the specific heat per nucleon at constant density,
$c_{\rho}(T)\equiv (\partial \varepsilon/\partial T)_{\rho}/\rho~$,
presented in Fig. 4.
A finite discontinuity of $c_{\rho}(T)$ arises
at the boundary between the mixed phase and  the gaseous phase.
This finite discontinuity
is caused by the fact that
$\lambda(T)=0$, but
$(\partial\lambda/\partial T)_{\rho} \neq 0$
at this boundary 
(see Fig. 2).

The negative values of the specific heat shown in Fig. 4
appear due to the parameterization of the surface tension (\ref{Sig}):   
its second derivative with respect to  temperature generates
a negative contribution in the vicinity of 
$ T_c $  for all densities. 
This feature of the model  is unphysical and it has to be 
modified. 

It should be emphasized that the energy density is continuous
at the boundary of the mixed phase and the gaseous phase, hence
the sharpness of the
peak in $c_{\rho}$ is entirely due to the strong temperature
dependence
of $\lambda(T)$ near this boundary. 
Moreover, at any $\rho < \rho_{\rm o}$
the maximum value of $c_{\rho}$ remains finite
and the peak width in $c_{\rho}(T)$ is nonzero in the thermodynamic
limit considered in our study. 
This is in contradiction with the expectation of Refs. \cite{Gu:98,Gu:99}
that an infinite peak of zero width will appear in $c_{\rho}(T)$ in this
limit.
Also a comment about the so-called ``boiling point''
is appropriate here.
This is a discontinuity in the energy density $\varepsilon(T)$ 
or, equivalently, a plateau in the
temperature as a function of the excitation energy. 
Our analysis shows that this type of behavior indeed happens 
at constant pressure, but not at constant density! This is similar to
the usual picture of a liquid-gas phase transition.
In Refs. \cite{Gu:98,Gu:99} a rapid  
increase of the energy density as a function of temperature
at fixed $\rho$ near the boundary of the mixed and gaseous phases
(see Fig.~3)
was misinterpreted as a manifestation of the ``boiling point''.

The results presented in Figs.~1-4 are obtained for $\tau=0$.
New possibilities appear
at non-zero values of the parameter $\tau$.
At $0<\tau \le 5/2$ the qualitative picture remains the same
as discussed above, although there are some
quantitative changes.
For $\tau > 5/2$  the condition ${\cal F}(s,T,\mu) < \infty$ 
is also satisfied
in the singularity point $s_l(T, \mu)$ 
for all  $T>T_c$ where $\sigma(T)=0$.
Therefore, the liquid-gas phase transition
extends now to all temperatures. Its properties
are, however, different for $\tau >7/2$ and for $\tau <7/2$
(see Fig.~5).
If $\tau >7/2$ the
gas density is always lower than $1/b$ as $\rho_{id}$ is finite.
Therefore, the
liquid-gas transition at $T>T_c$ remains
the 1-st order phase transition with discontinuities
of baryonic density, entropy and energy densities.

At $5/2 < \tau < 7/2$ the baryonic density of the gas
in the mixed phase, 
$\rho_g^{mix} \equiv \rho_{id}^{mix}(T)/( 1+ b~\rho_{id}^{mix}(T) )$,
becomes equal to that of the liquid at $T>T_c$, i.e.,  
$\rho_g^{mix}=1/b\equiv \rho_{\rm o}$, since
\begin{eqnarray}
\rho_{id}^{mix}(T)~ \equiv~\rho_{id}(T,\mu^*(T))~
& = &  \left( \frac{ mT }{2 \pi}\right)^{3/2}
\left[z_1 \exp\left(-~\frac{W}{T}\right)\right.
\nonumber \\
&+& \sum_{k=2}^{\infty}\left.
k^{\frac{5}{2}- \tau}
\exp\left(-~ \frac{\sigma ~k^{2/3}}{T}\right)\right]~ \rightarrow \infty\,\,,
\label{rhoidmix}
\end{eqnarray}
if surface tension vanishes $\sigma = 0$.
It is easy to prove that the entropy and energy densities
for the liquid and gas phases are also equal to each other.
There are discontinuities only in the derivatives of these densities
over $T$ and $\mu$, i.e., $p(T,\mu)$  has discontinuities
of its second derivatives.
Therefore,  the liquid-gas transition at $T>T_c$ for $5/2 < \tau < 7/2$
becomes the 2-nd order phase transition.
According to standard definition, the point $T=T_c$,
$\rho = 1/b$ separating the first and second order
transitions is the tricritical point.
One can see that this point is now at a finite pressure.

It is interesting to note that at $\tau >0$ the mixed
phase boundary shown in Fig.5 is not so steep function of $T$ as
in the case $\tau=0$ presented in Fig.1. Therefore, the peak in the specific
heat discussed above becomes less pronounced.

\vspace{0.2cm}
In conclusion, the simplified version of the SMM
is solved analytically
in the grand canonical ensemble.
The progress is achieved by 
reducing the description of phase transitions
to the investigation  of the isobaric
partition function singularities. The model clearly 
demonstrates a 1-st order
phase transition of the liquid-gas type.
The considered system has  peculiar
properties near the boundary of the mixed and gaseous
phases.
The rapid change
of the thermodynamical functions with $T$ at fixed $\rho$ 
takes place 
near this  boundary
due to the disappearance of the infinite liquid fragment. 
This leads to leveling of 
the caloric curves shown in Fig.~3
at temperatures between 6 -- 10 MeV depending on the density. 
As a consequence 
a sharp peak and a finite discontinuity
are developed 
in the specific heat $c_{\rho}(T)$ 
at the boundary of the mixed and gaseous phases. 

The phase diagram appears to be rather
sensitive to the value of the parameter $\tau$
in the Fisher's free energy term included in our treatment.
New interesting possibilities for the phase diagram emerge
for $\tau >5/2$  in comparison
with the case when $\tau<5/2$.
The case $5/2<\tau<7/2$ is particularly interesting
because of the appearance of the tricritical point separating the
1-st and 2-nd order phase transitions.

\vspace*{0.2cm}

\begin{center}
{\bf  Acknowledgments.}  
\end{center}

The authors 
are grateful to 
D. Blaschke, J. Polonyi 
and P.T.~Reuter for stimulating discussions. 
We  thank the Alexander von Humboldt Foundation
and DAAD (Germany) for the financial support. 
The research described in this publication was made possible in part by
Award No. UP1-2119 of the U.S. Civilian Research \& Development
Foundation for the Independent States of the Former Soviet Union
(CRDF).
K.A.B. gratefully acknowledges
the warm hospitality of the Particle- and Astrophysics Group
of the University of Rostock, where this talk was given.

\input capts.tex

\end{document}

%% file: capts.tex

\clearpage

\begin{figure}
\begin{center}
\mbox{\psfig{figure=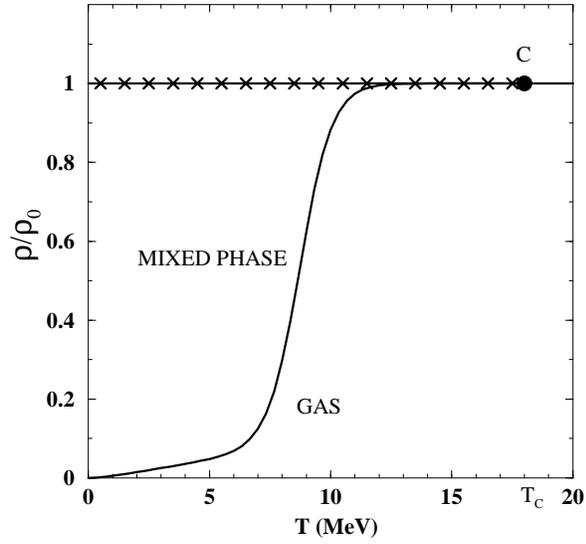,width=12.cm}}
\end{center}

\vspace*{-0.5cm}

\caption{\label{fig:one}
Phase diagram in the $(T,\rho)$-plane.
The mixed phase and pure gaseous phase boundary
is shown by the solid line.
The pure liquid phase (shown by crosses) corresponds
to
the fixed density $\rho = \rho_{\rm o}$.
Point $C$ is the critical point,
at $T>T_c$ only the pure gaseous phase
exists.
}
\end{figure}


\vspace*{-0.5cm}

\begin{figure}   
\begin{center}
\mbox{\psfig{figure=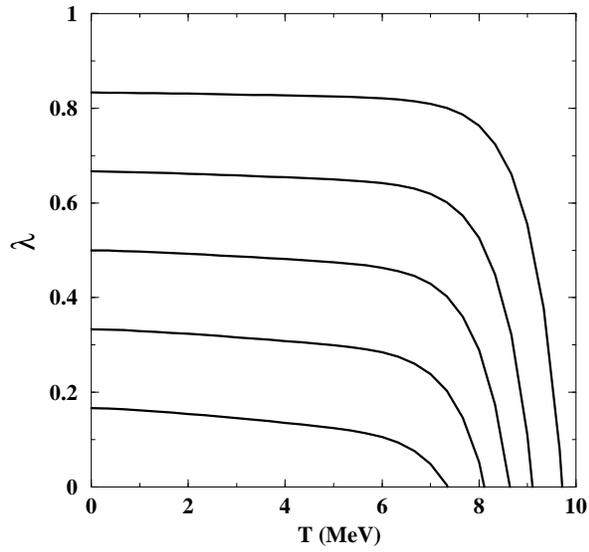,width=12.cm}}
\end{center}

\vspace*{-0.5cm}

\caption{\label{fig:two}
Volume fraction $\lambda(T)$ of the liquid
inside the mixed phase is
shown as a function of temperature
for fixed nucleon densities ${\rho}/{\rho_{\rm o}} = 1/6, 1/3, 1/2, 2/3,
5/6$
(from bottom to top).
}
\end{figure}

\newpage
\clearpage

\begin{figure}
\begin{center}
\mbox{\psfig{figure=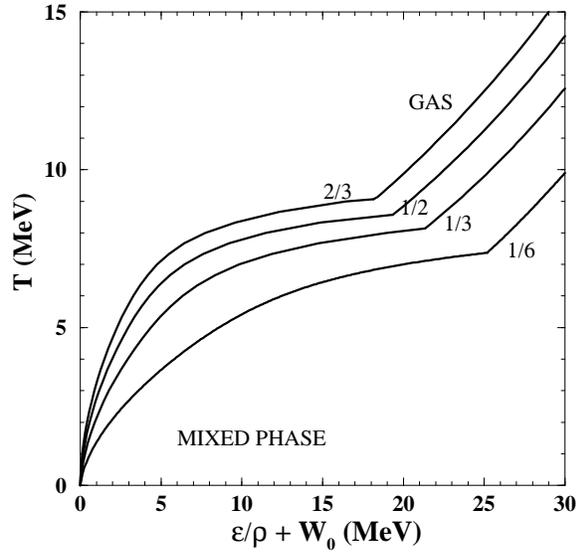,width=12.cm}}
\end{center}

\vspace*{-0.5cm}

\caption{\label{fig:three}
Temperature as a function of energy density per nucleon
(caloric curve)
is shown for fixed nucleon densities ${\rho}/{\rho_{\rm o}} = 1/6, 1/3,
1/2, 2/3$.
}
\end{figure}


\begin{figure}
\begin{center}
\mbox{\psfig{figure=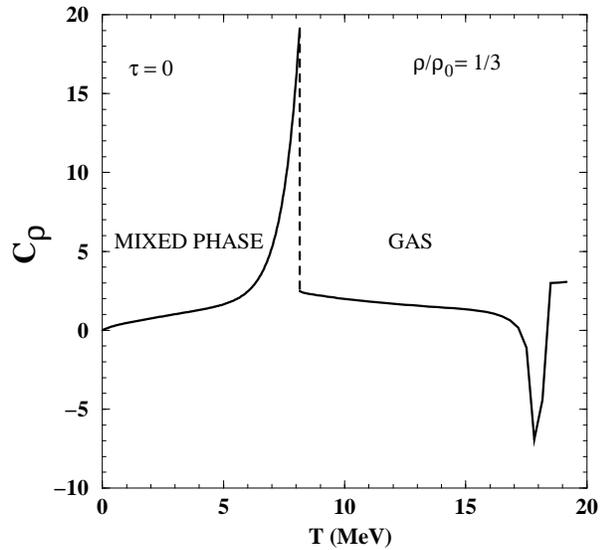,width=12.cm}}
\end{center}

\vspace*{-0.5cm}

\caption{\label{fig:four}
Specific heat per nucleon as a function of temperature
at fixed nucleon density ${\rho}/{\rho_{\rm o}} = 1/3$. The dashed line
shows the finite discontinuity of $c_{\rho}(T)$
at the boundary of the mixed and gaseous phases.
The negative specific heat appears in the vicinity of $T = 18$ MeV. 
}
\end{figure}

\newpage
\clearpage

\begin{figure}
\begin{center}
\mbox{\psfig{figure=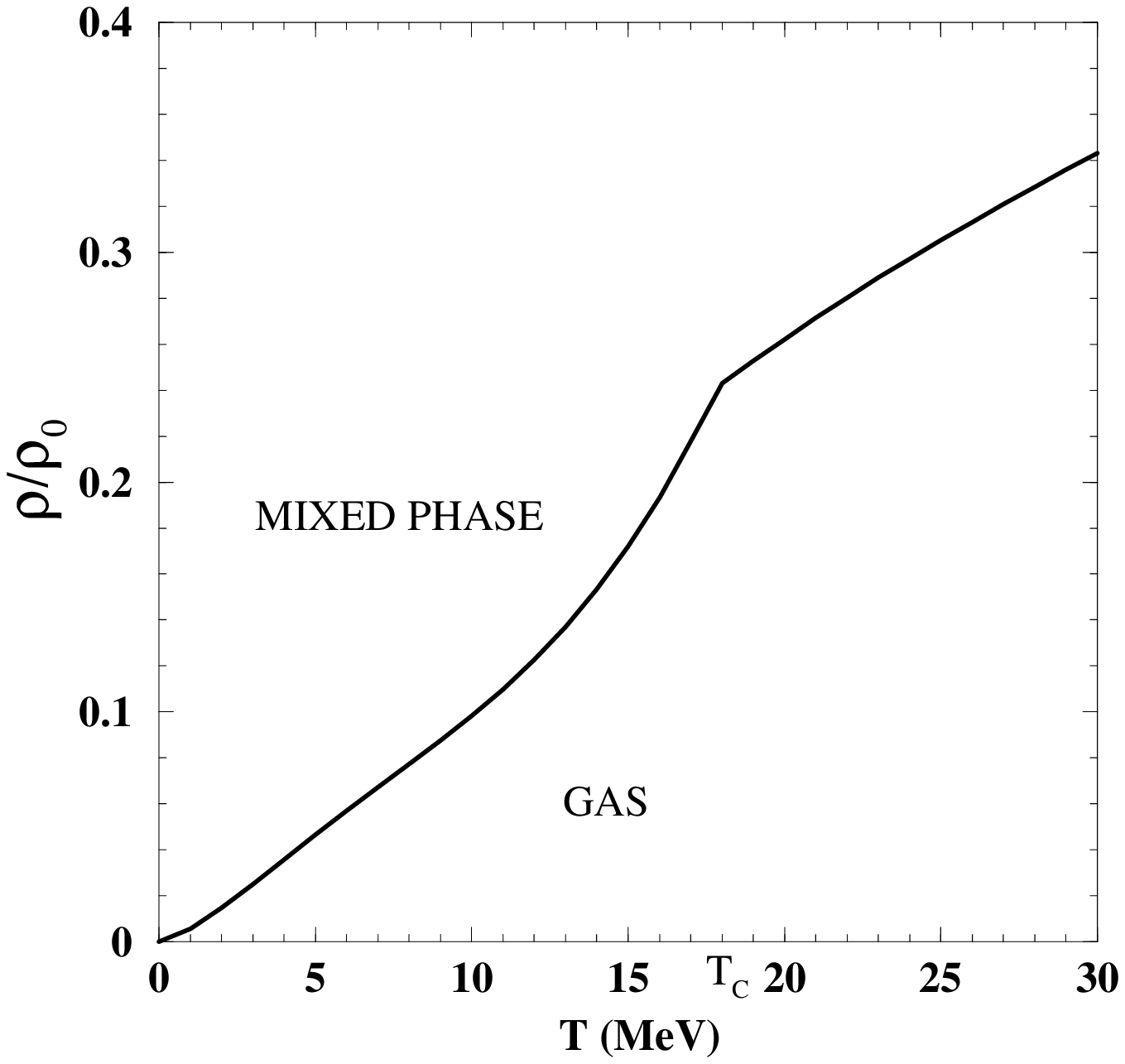,width=12.0cm}}

\vspace*{ 1.2cm}

\mbox{\psfig{figure=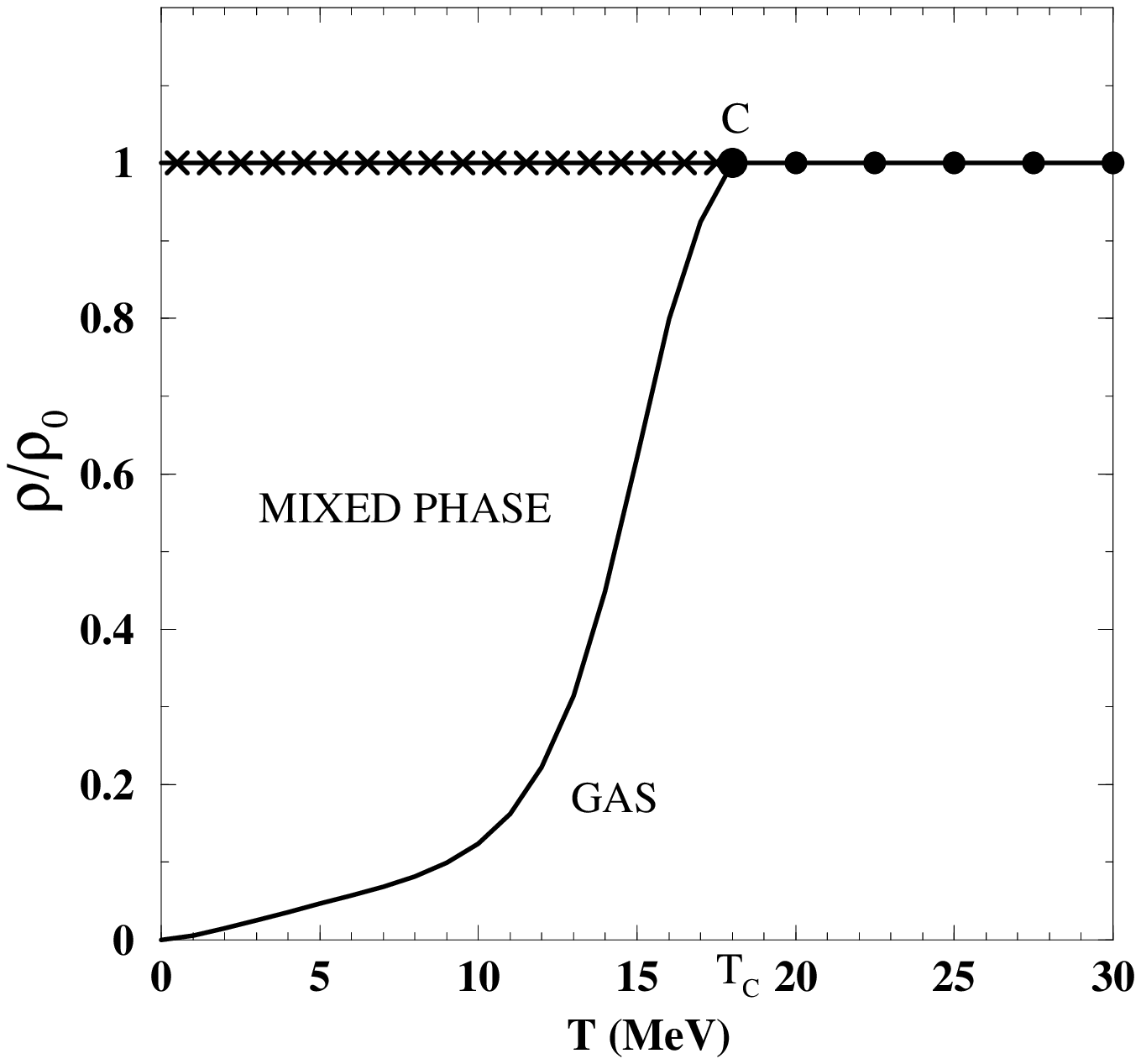,width=12.0cm}}
\end{center}

\vspace*{-0.5cm}

\noindent
\caption{\label{fig:five}
Phase diagrams in
$T-\rho$ plane for $\tau = 3.6$ (upper panel) and $\tau = 2.6$ (lower panel).
Point $C$ in the lower panel is the tricritical point.   
Crosses correspond to the liquid phase of the first order phase transition
and dots correspond to the states of the second order one.
}
\end{figure}

%% file: main.bbl
\begin{thebibliography}{99}

\bibitem{Bo:95}
J.P. Bondorf, A.S. Botvina, A.S. Iljinov, I.N. Mishustin, K.S.~Sneppen,
Phys. Rep. {\bf 257} (1995) 131.

\bibitem{Gr:97} 
D.H.E. Gross, Phys. Rep. {\bf 279} (1997) 119.

\bibitem{Ch:95} 
K.C. Chase and A.Z. Mekjian, Phys. Rev. {\bf C 52}
(1995) R2339. 

\bibitem{Gu:98}
S. Das Gupta and A.Z. Mekjian, Phys. Rev. {\bf C 57}
(1998) 1361.

\bibitem{Gu:99}
S. Das Gupta, A. Majumder, S. Pratt and A. Mekjian,
nucl-th/9903007 (1999).

\bibitem{Gu:00}
S. Das Gupta, A.Z. Mekjian and M.B. Tsang,
nucl-th/0009033 (2000).

\bibitem{Bu:00}
 K.A. Bugaev, M.I. Gorenstein, I.N.~Mishustin and W.~Greiner,
Phys. Rev. {\bf C62} (2000) 044320.

\bibitem{Go:81}
M.I. Gorenstein, V.K. Petrov and G.M. Zinovjev, Phys. Lett.
{\bf B 106} (1981) 327; \\
M.I. Gorenstein, W. Greiner and S.N. Yang,
J. Phys. {\bf G 24} (1998) 725.

\bibitem{Bu:00B}
K.A. Bugaev, M.I. Gorenstein, I.N. Mishustin  and  W. Greiner,
{\bf nucl-th/0007062} (2000) 3 p.

\bibitem{Bu:00C}
K.A. Bugaev, M.I. Gorenstein, I.N. Mishustin  and  W. Greiner,
{\bf nucl-th/0011055} (2000) 7 p.



\bibitem{Fisher:67}
M.E. Fisher, Physics {\bf 3} (1967) 255. 


\end{thebibliography}
